\documentstyle{article}
\date{\today}

\begin{document}
\title{QCD Sum Rules: Isospin Symmetry Breakings in Pion-Nucleon 
Couplings}

\author{{W-Y. P. Hwang,$^{1,2}$ Ze-sen Yang$^3$, Y.S. Zhong,$^3$ 
Z. N. Zhou,$^3$ and Shi-lin Zhu$^3$}\\
{$^1$Department of Physics, National Taiwan University, Taipei, 
Taiwan 10764}\\
{$^2$Center for Theoretical Physics, Laboratory for Nuclear Science 
       and}\\
{Department of Physics, Massachusetts Institute of Technology, }\\ 
{Cambridge, Massachusetts 02139}\\
{$^3$Department of Physics, Peking University, Beijing 100871, China}
}
\maketitle

\begin{center}
\begin{minipage}{120mm}
\vskip 0.6in
\begin{center}{\bf Abstract}\end{center}
{\large
We use the method of QCD sum rules in the presence of an external 
pion field to investigate isospin symmetry breakings in pion-nucleon
couplings. Possible manifestations of such isospin symmetry breaking are 
examined in the context of low-energy nucleon-nucleon ($NN$) scattering.
We discuss numerical results in relation to both the existing data and 
other theoretical predictions.
\vskip 0.5 true cm
PACS Indices: 24.80.Dc; 21.30.+y; 13.75.Cs
}
\end{minipage}
\end{center}

\large

\vskip 1.5cm

\section*{I. Introduction}

\par
Isospin symmetry and its possible violations constitute an integral part 
of nuclear physics \cite{Henley}, not only in the sense that good symmetry 
offers an effective classification of nuclear or hadronic structure but 
also in the proximity of the isospin symmetry concept to other fundamental 
ideas such as chiral symmetry. In general, isospin symmetry breaking 
effects observed in nuclear physics may arise 
from different sources such as, in addition to the well known 
electromagnetic interactions, strong interactions but augmented with 
hadron mass differences (such as the $n-p$ or $\pi^\pm-\pi^0$ mass 
difference) or with different meson-nucleon couplings (such as non-universal 
$\pi NN$ couplings) or with isospin mixings (such as $\rho-\omega$ or 
$\pi-\eta$ mixing). Determination of isospin symmetry breaking effects
requires information on quantities which may not be directly accessible 
from experiments, such as the meson-nucleon couplings or the isospin 
mixing parameters. Accordingly, reliable theoretical predictions become 
indispensible for advancing our understanding of the problem.

Unfortunately, it appears that a quantitative treatment of isospin symmetry 
breakings may involve the capability of treating strong 
interactions at a quantitative level or, in other words, the ability
to handle the complications due to quantum chromodynamics (QCD) in 
a reliable manner. To some extent, the method of QCD sum rules, as proposed
originally by Shifman, Vainshtein, and Zakharov \cite{SVZ} and adopted,
or extended, by many others \cite{RRY,IOFFE,BALIT}, 
incorporates the nonperturbative QCD effects through various condensates 
(associated with the nontrivial QCD vacuum), thereby offering some hope 
of being able to treat strong interactions in a quantitative manner.
\par
In this work, we wish to focus on the determination of isospin symmetry
breakings in pion-nucleon couplings, with some emphasis on possible
manifestations of such symmetry breaking in nucleon-nucleon 
scattering. As the present effort is based
primarily on the method of QCD sum rules, it differs much from the 
quark-model calculations \cite{Cao,Gardner}; as we shall see, this
work is considerably more extensive than
a recent work \cite{Meissner} also making use of the QCD sum rule method. 
Specifically, we wish to adopt a method in which
the pion field $\pi$ is treated as the 
external field. The idea of treating the pion as an external field was 
first suggested in \cite{IOFFE} and has not been used in any extensive
manner until recently \cite{PST, Hwang}.

\vskip 1.5 true cm
\section*{II. Isospin Symmetry Breakings in Pion-Nucleon Couplings}
\bigskip

In the method of QCD sum rules in the presence of an external field, we 
attempt to evaluate, both at the quark and hadron levels, the 
correlation function specified by
\begin{equation}
\Pi(p)\equiv i\int d^4x e^{ip\cdot x} \langle 0 |T(\eta_p(x) {\bar\eta}_p(0)
|0\rangle \mid_\pi,
\end{equation}
where we have
\begin{equation}
\begin{array}{ll}
\eta_p(x)&= \epsilon^{abc}\{u^{aT}(x)C\gamma_\mu u^b(x)\}\gamma^5\gamma^\mu
            d^c(x) \\
{\bar\eta}_p(y) &= \epsilon^{abc}\{{\bar u}^b(y) \gamma_\nu C 
         {\bar u}^{aT}(y) \} {\bar d}^c(y) \gamma^\nu \gamma^5.
\end{array}
\end{equation}
By evaluating the appropriate correlation function in the presence of the
external pion field, we may determine the pion-nucleon couplings 
$g_{\pi^0 pp}$, $g_{\pi^0 nn}$, and $g_{\pi^+pn}$. An alternative method
is to evaluate correlation funtions with $T$ product of currents sandwiched 
between the vacuum and the one-pion state, as first suggested by \cite{Crai}
and recently adopted by some authors \cite{Birse}. It appears that the 
induced condensates involved in the external field method are now well 
understood and it becomes relatively straightforward to perform calculations
to higher dimensions (as required for reliable predictions). This is what we 
choose to use in this paper and we use it to obtain isospin symmetry 
breakings in $\pi NN$ couplings. 

The above correlation function allows us to determine the $\pi^0 pp $ 
coupling. At the quark level, we have
\begin{equation}
\langle 0| T\eta_p(x) {\bar \eta}_p(0) |0\rangle_{\pi^0} = 2 {\rm i} 
\epsilon^{abc}
\epsilon^{a^{\prime} b^{\prime} c^{\prime}} {\bf Tr} \{ S_u^{b b^{\prime}}(x)
\gamma_{\nu} C [S_u^{a a^{\prime}}(x) ]^T C \gamma_{\mu} \}\,
\gamma_5 \gamma_{\mu} S_d^{c c^{\prime}}(x) \gamma_{\nu} \gamma_5.
\end{equation}
Here the quark propagator is given by
\begin{equation}
S(x)=S^{(0)}(x)+S^{(\pi)}(x),
\end{equation}
with
\begin{equation}
\begin{array}{ll}
{\rm i}S^{(0)ab}(x) =&
 {{\rm i}\delta^{ab}\over 2\pi^2 x^4}{\hat x}
   +{{\rm i}\over 32\pi^2 x^2}{\lambda^n_{ab}\over 2}g_c G_{\mu\nu}^n 
    (\sigma^{\mu\nu}{\hat x}+{\hat x}\sigma^{\mu\nu})
   -{\delta^{ab}\over 12}\langle {\bar q} q\rangle         \\
 &+{\delta^{ab} x^2\over 192} \langle g_c {\bar q}\sigma \cdot G q\rangle
 -{m_q \delta^{ab}\over 4\pi^2 x^2}
 +{m_q\over 32\pi^2}{\lambda_{ab}^n\over 2} G^n_{\mu\nu} \sigma^{\mu\nu} 
 \ln (-x^2) \\
 &-{\delta^{ab} \langle g^2_c G^2 \rangle \over 2^9 \times 3\pi^2} 
 m_q x^2 \ln (-x^2) 
 +{i \delta^{ab} m_q\over 48} \langle {\bar q} q\rangle {\hat x}
 -{1\over 2^7\times 3^2} i m_q \langle g_c {\bar q}\sigma \cdot G q\rangle 
 \delta^{ab} x^2  \, ,
\end{array}
\end{equation}
and
\begin{equation}
\begin{array}{ll}
{\rm i}S^{(\pi)ab}(x) =&-{{\rm i}\delta^{ab}\over 4\pi^2 x^2}
    g_q {\vec \tau}\cdot {\vec \pi} \gamma_5 
   +{{\rm i}\delta^{ab}\over 24}g_q {\vec \tau}\cdot {\vec \pi} 
   \gamma_5 \chi \langle {\bar q}q \rangle \\
&  -{{\rm i}\delta^{ab}\over 384}m_0^\pi g_q {\vec \tau}\cdot {\vec \pi} 
\gamma_5 \langle {\bar q}q\rangle x^2  
  +{\rm i\over 2^5 \pi^2} g_c g_q {\vec \tau}\cdot {\vec \pi} \gamma_5
  {\lambda^n_{ab}\over 2} G^n_{\mu\nu} \sigma^{\mu\nu}\ln (-x^2) \\
& -{{\rm i\delta^{ab}}\over 2^9\times 3 \pi^2}  g_q {\vec \tau}\cdot {\vec \pi} \gamma_5 
   \langle g^2_c G^2 \rangle x^2 \ln (-x^2)
  -{\delta^{ab}\over 48}g_q {\vec \tau}\cdot {\vec \pi} 
  \gamma_5 \langle {\bar q}q\rangle {\hat x} \, ,
\end{array}
\end{equation}
where we have introduced ${\hat x} \equiv x_\mu \gamma^\mu$, $<{\bar 
\psi} (0)i \gamma^5 \tau^j \psi(0)> \equiv g_q \chi \pi^j <{\bar q}q>$, 
and $<{\bar \psi}(0) i \gamma^5 \tau^j
g_c \sigma\cdot G\psi(0)> \equiv g_q m_0^\pi \tau^j <{\bar q}q>$ with 
$\psi$ the isospin doublet consisting of $u$ and $d$ quarks. 
The various terms in the quark propogators $iS^{(0)ab}(x)$ and 
$iS^{(\pi)ab}(x)$ may be represented pictorially by the diagrams shown in 
Figs. 1 and 2, respectively. We wish to stress that we have in fact performed
our calculations in momentum space especially when ambiguities arise and,
in addition, special attention has been directed to adoption of the
$\gamma_5$ in $d-$dimensions in relation to dimensional regularization.

To incorporate isospin symmetry violations, we note that $<{\bar q} q>$ 
appearing in the quark propagator needs to be interpreted properly; in 
particular, it is to be understood as $<{\bar u}u>$ in the $(uu)$ channel, 
or as $<{\bar d} d>$ in the $(dd)$ channel, or as $(<{\bar u}u> 
+<{\bar d}d>)/2$ for the $(d\to u)$ or $(u\to d)$ propagation. We also
note that Fig. 2(f) has been drawn in a way to indicate that the choice 
of $q$ in $<{\bar q} q>$ should follow the flavor of the quark connecting
to $x=0$ (on the right hand side of the diagram) $-$ a result following 
usage of the 
equation of motion. No potential ambiguity should arise for all the other
diagrams shown in Figs. 1 and 2. The assumption has also been 
made that, once we have made the suitable interpretation of $<{\bar q} q>$, 
$\chi$ and $m_0^\pi$ are considered to be universal. In other words, we
refrain from introducing, at the quark level, further unknown sources for 
isospin symmetry violations; our assumption regarding the universal
nature of the susceptibilities $\chi$ and $m_0^\pi$ is further supported by
the derivation of induced condensates as illustrated in \cite{Hwang}. This
is a crucial point since the terms in $\chi <{\bar u} u>$ and $\chi 
{\bar d} d>$ turn out to be one of the most important contributions
which distinguish the coupling $g_{\pi^0 pp}$ from $g_{\pi^0 nn}$.

We note that, for the determination of $\pi^- pn$ or $\pi^+ np$ coupling, 
we find
\begin{equation}
\langle 0| T\eta_p(x) {\bar \eta}_n(0) |0\rangle_{\pi^-} =-4{\rm i} \epsilon^{abc}
\epsilon^{a^{\prime} b^{\prime} c^{\prime}}  \gamma_{\mu} \gamma_5 S_d^{a a^{\prime}}(x)
\gamma_{\nu} C [S_{\phi}^{b b^{\prime}}(x) ]^T C \gamma_{\mu}
S_u^{c c^{\prime}}(x)\gamma_5 \gamma_{\nu}.
\end{equation}
We note the absence of ${\rm i}S^{(0)}(x)$ in the present case, due to a 
change in the electric charge. We also note that the structure (which
does not involve the trace of some product) seems quite different 
from the $\pi^0 pp$ case, but 
we will see that isospin symmetry is indeed there should we assume 
$m_u = m_d$ and $<{\bar u} u> = <{\bar d} d>$. Finally, to obtain QCD
sum rules up to dimension eight (as compared to the leading perturbative 
term), we need to enumerate diagrams which are proportional to
$ 1$, $\chi \langle {\bar q} q \rangle $, $\langle g^2_c G^2 \rangle$,
$m^{\pi}_0 \langle {\bar q} q \rangle $, $\langle {\bar q} q \rangle^2$,
$\langle {\bar q} q \rangle \langle g_c {\bar q} \sigma \cdot G q \rangle$,
$m_q \langle {\bar q} q \rangle$,
and $m_q \langle {\bar q} g_c \sigma \cdot G q \rangle$. Here it is clearly 
of importance to be able to distinguish between $<{\bar d} {\hat O} d>$ and 
$<{\bar u} {\hat O} u>$ in the quark propagators used to evaluate the
correlation functions.

On the other hand, we may parametrize the correlation functions at the
hadronic level in the standard manner.
\begin{equation}
\int d^4x \langle 0| T\eta_p(x) {\bar \eta}_n(0) |0\rangle\mid_{\pi^-} =
\lambda_n \lambda_p
\frac{\rm i}{{\hat p}-m_p} (-\gamma_5 \sqrt{2} g_{\pi^- pn}) 
\frac{\rm i}{{\hat p}-m_n} +\cdots,
\end{equation}
\begin{equation}
\int d^4x \langle 0| T\eta_p(x) {\bar \eta}_p(0) |0\rangle\mid_{\pi^0} =
\lambda_p^2 \frac{\rm i}{{\hat p}-m_p} + \lambda_p^2
\frac{\rm i}{{\hat p}-m_p} (-\gamma_5 g_{\pi^0 pp} )
\frac{\rm i}{{\hat p}-m_p}+\cdots.
\end{equation}
We consider the various diagrams shown in Fig. 3 and obtain the 
quark-level expression (referred to as ``l.h.s.'') for the $\pi^0 pp$ 
coupling (directly from the momentum-space calculations), 
\begin{equation}
\begin{array}{ll}
\frac{1}{(2\pi)^4} g_{\pi^0 dd} \phi \gamma_5 \{
\frac{p^4 \ln(-p^2)}{2} -{\chi\over 2} a_d p^2\ln(-p^2) 
 -\frac{11}{24}b\ln(-p^2) &\\
 +\frac{4}{3}a_u^2 \frac{1}{p^2}
 -\frac{1}{3}a_u^2 m^2_0 \frac{1}{p^4}
 +\frac{\ln(-p^2) +\gamma_E}{p^2}m^2_0 [\frac{7}{8}
   (a_u m_d+a_d m_u)-\frac{1}{4}m_u a_u ]   &\\
 +\frac{m_0^2}{p^2}[\frac{5}{12}m_u a_u-\frac{1}{8}m_d a_d
  +\frac{67}{48}m_d a_u +\frac{115}{48}m_u a_d ]\} &
\end{array}
\end{equation}
with $a_q \equiv -(2\pi)^2<{\bar q}q>$, $<{\bar q} \sigma\cdot G> \equiv
- m_0^2 <{\bar q} q>$ and $b \equiv <g_c^2 G^2>$. 
Accordingly, we obtain, upon Borel transform, the following QCD sum rule 
for the $\pi^0 pp$ coupling:
\begin{equation}
\begin{array}{ll}
& M^6  -{\chi\over 2} a_d M^4  -{11\over 24} b M^2  
 +{4\over 3} a_u^2  + {m_0^2\over 3}a_u^2 {1\over M^2} \\
& + m_0^2 \ln {M^2\over \mu^2} 
  [\frac{7}{8}(a_u m_d+a_d m_u)-\frac{1}{4}m_u a_u ]  \\
& +m_0^2 [\frac{5}{12}m_u a_u-\frac{1}{8}m_d a_d
  +\frac{67}{48}m_d a_u +\frac{115}{48}m_u a_d ] \\
=&\frac{g_{\pi^0 pp}}{g_{\pi^0 dd}} {\tilde \lambda}^2_p 
e^{-\frac{m_p^2}{M^2}} +\mbox{excited states}+ \mbox{continuum},
\end{array}
\end{equation}
where $M$ is the Borel mass and ${\tilde \lambda}_p^2 = (2\pi)^4 \lambda_p^2$.
A similar expression can be obtained for the $\pi^0 nn$ coupling through 
the replacement $u\leftrightarrow d$. It is of some interest to note that
the contribution from Fig. 3(d-1) is canceled by that from Fig. 3(d-2), 
so that the induced susceptibility $m_0^\pi$ does not enter in the
determination of strong $\pi NN$ coupling, an aspect already observed in
\cite{PST}.

For the $\pi^+ np$ coupling, we have, with ${\bar a}\equiv (a_u+a_d)$,
\begin{equation}
\begin{array}{ll}
& M^6  -{\chi\over 2} {\bar a} M^4  -{11\over 24} b M^2  
 +{4\over 3} a_u a_d + {1\over 3}(a_d -a_u) a_d -(m_u-m_d) a_u M^2\\
& + m_0^2 \{ -{1\over 2} a_u m_u -{5 \over 24} a_d m_d 
+{29\over 12} a_u m_d + {19 \over 8} a_d m_u \}\\
& + m_0^2 \ln {M^2\over \mu^2} \{ a_u m_d +a_d m_u - {1\over 4} m_u 
  (a_d +a_u) \} + {m_0^2\over 3}a_u a_d {1\over M^2} \\
=&\frac{g_{\pi^+ np}}{g_{\pi^+ du}} (2\pi)^4 \lambda_n \lambda_p 
e^{-\frac{{\bar m}^2}{M^2}} +\mbox{excited states}+\mbox{continuum}.
\end{array}
\end{equation}
As a standard practice, we may improve the applicability of these QCD 
sum rules by (1) taking into account the anomalous dimension of each term
(operator) and (2) invoking the continuum approximation in which contributions
from excited states and the continuum are approximated by what we may 
obtain at the quark level above a certain threshold $W^2$. Results from
such improvements can be duplicated easily from Yang et al. \cite{Yang}.
In this way, we obtain, for the $\pi^0 pp$ coupling \cite{Yang,PST},
\begin{equation}
\begin{array}{ll}
& M^6 L^{-4/9} E_2  -{\chi\over 2} a_d M^4 L^{2/9} E_1  
-{11\over 24} b M^2 E_0  +{4\over 3} a_u^2 L^{4/9}\\ 
& + {m_0^2\over 3}a_u^2 {1\over M^2} L^{-2/27}
  + m_0^2 \ln {M^2\over \mu^2}\, L^{-26/27}
  [\frac{7}{8}(a_u m_d+a_d m_u)-\frac{1}{4}m_u a_u ]  \\
& +m_0^2 L^{-26/27} [\frac{5}{12}m_u a_u-\frac{1}{8}m_d a_d
  +\frac{67}{48}m_d a_u +\frac{115}{48}m_u a_d ] \\
=&\frac{g_{\pi^0 pp}}{g_{\pi^0 dd}} {\tilde \lambda}^2_p 
e^{-\frac{m_p^2}{M^2}},
\end{array}
\end{equation}
and, for the $\pi^+ np$ coupling,
\begin{equation}
\begin{array}{ll}
& M^6 L^{-4/9} E_2 -{\chi\over 2} {\bar a} M^4 L^{2/9} 
-{11\over 24} b M^2 E_0  +\{ {4\over 3} a_u a_d + {1\over 3}(a_d -a_u) 
a_d \} L^{4/9} \\ 
& - (m_u-m_d) a_u M^2 L^{-4/9} E_0
  + m_0^2 L^{-26/27} \{ -{1\over 2} a_u m_u -{5 \over 24} a_d m_d 
+{29\over 12} a_u m_d + {19 \over 8} a_d m_u \}\\
& + m_0^2 \ln {M^2\over \mu^2}\, L^{-26/27} \{ a_u m_d +a_d m_u 
- {1\over 4} m_u  (a_d +a_u) \} + {m_0^2\over 3}a_u a_d L^{-2/27}
{1\over M^2} \\
=&\frac{g_{\pi^+ np}}{g_{\pi^+ du}} (2\pi)^4 \lambda_n \lambda_p 
e^{-\frac{{\bar m}^2}{M^2}}.
\end{array}
\end{equation}
Here $E_0 = 1 - e^{-x}$, $E_1 = 1 -(1+x)e^{-x}$, and $E_2 = 1 - (1+x+
{x^2\over 2})e^{-x}$ with $x=W^2/M^2$, and $L=0.621 ln(10M)$. 

There are serveral sources for isospin symmetry breakings, viz.:

At the hadronic level (r.h.s.), we have $m_n \not= m_p$, $\lambda_n \not= 
\lambda_p$, and $W^2_n \not= W^2_p$ (with $W^2$ the threshold parameter in 
the continuum approximation for treating the excited states and the 
continuum). Apart from the well known neutron-proton mass difference, we 
use the parameters previously adopted by Yang et al. \cite{Yang}. 

At the quark level (l.h.s.), the isospin symmetry breakings caused by 
$<{\bar d}d> \not= <{\bar u}u>$ and $m_d \not= m_u$ have been made explicit
in the formulae given above. Such effects have also been considered by 
other authors \cite{Meissner, Yang}. 

In addition, there is 
an effect due to the non-universal pion-quark couplings. Conceptually,
we may start with a universal pion-quark coupling, such as in the 
effective chiral quark theory, but the vertex renormalizations due to
electromagnetic interactions, as shown in Fig. 4 and also pointed out by
Cao and Hwang \cite{Cao}, modify $\pi^0 uu$, $\pi^0 dd$, and $\pi^+ud$
in a different manner. Here we have adopted the dimensional regularization
scheme to work out such differences and conclude that the non-universal
pion-quark couplings so obtained could be a reasonably important source for 
isospin symmetry breaking. Specifically, we obtain the following expressions
for the pion-quark couplings. 
\begin{equation}
\begin{array}{ll}
g_{\pi^+du} &\longrightarrow g_{\pi^+du}\{1+{\alpha\over 4\pi}(-{43\over 18}
+{1\over 6} \gamma_E)\},\\
g_{\pi^0uu} &\longrightarrow g_{\pi^0uu}\{1+{\alpha\over 4\pi}({52\over 9}
-{4\over 3} \gamma_E)\},\\
g_{\pi^0dd} &\longrightarrow g_{\pi^0dd}\{1+{\alpha\over 4\pi}({13\over 9}
-{1\over 3} \gamma_E)\}.\\
\end{array}
\end{equation}
Numerically, these expressions give rise to $2.18 \times 10^{-3}$ to 
the charge symmetry 
breaking term $g_{\pi^0 uu} -g_{\pi^0 dd}$ and $4.2 \times 10^{-3}$ for
the charge dependent combination $g_{\pi^0 uu} - g_{\pi^+ du}$.

To perform numerical analyses, we may introduce the following parameters: 
\begin{eqnarray*}
\lambda \equiv &\frac{\lambda_p +\lambda_n}{2}, \qquad
\sigma \equiv \frac{\lambda_n -\lambda_p}{\lambda}; \\
{\bar m} \equiv &\frac{m_p +m_n}{2}, \qquad
\epsilon \equiv \frac{m_n -m_p}{\bar m}.
\end{eqnarray*}
\begin{eqnarray*}
\frac{g_{\pi^0 pp}}{g_{\pi^0 dd}} \equiv &K\\
\frac{g_{\pi^0 nn}}{g_{\pi^0 uu}} \equiv &K (1+\rho_1)\\
\frac{g_{\pi^- pn}}{g_{\pi^- ud}} \equiv &K (1+\rho_2)\\
\frac{g_{\pi^+ np}}{g_{\pi^+ du}} \equiv &K (1+\rho_3)
\end{eqnarray*}
Moving the pion-quark coupling to the phenomenological side, we may write 
the r.h.s. as follows:
\begin{eqnarray*}
\pi^0 pp &\longrightarrow  K \lambda^2 e^{-\frac{m^2}{M^2}} 
(1 -\sigma +\epsilon \frac{m^2}{M^2} ),\\
\pi^0 nn &\longrightarrow  K \lambda^2 e^{-\frac{m^2}{M^2}} 
(1 +\sigma -\epsilon \frac{m^2}{M^2} +\rho_1 ),\\
\pi^- pn &\longrightarrow  K \lambda^2 e^{-\frac{m^2}{M^2}} (1  +\rho_2),\\
\pi^+ np &\longrightarrow  K \lambda^2 e^{-\frac{m^2}{M^2}} (1  +\rho_3 ).
\end{eqnarray*}
The various parameters which we adopt are $\sigma =-10^{-3}$ \cite{Yang}, 
$\epsilon =1.3 \times 10^{-3}$,
$\gamma =-6.57 \times 10^{-3}$ \cite{ChPT}, $a_q=0.546\, \mbox{GeV}^3$, 
$m_u =5.1\, MeV$, $m_d=8.9\, MeV$, $b=0.474\, \mbox{GeV}^4$, 
$m_0^2 =0.8\, \mbox{GeV}^2$, $\mu = 0.5\, GeV$, and $\chi a = -3.14\, 
GeV^2$ \cite{Hwang}.
The working interval for analyzing the QCD sum rules for the various  
$\pi NN$ couplings is $0.8\mbox{GeV}^2 \leq 
M_B^2\leq 1.5\mbox{GeV}^2$, a standard choice for analyzing the various 
QCD sum rules associated with the nucleon.

In this way, we find 
\begin{equation}
{{g_{\pi^0 nn}\over g_{\pi^0 uu}} - {g_{\pi^0 pp}\over g_{\pi^0 dd}}
 \over {g_{\pi^0 NN}\over g_{\pi^0 QQ}}\mid_{average} } = 0.58 \%
\end{equation}
\begin{equation}
{{g_{\pi^+ np}\over g_{\pi^+ du}} - {g_{\pi^0 pp}\over g_{\pi^0 dd}}
 \over {g_{\pi^0 NN}\over g_{\pi^0 QQ}}\mid_{average} } = 0.35 \%
\end{equation}
Combining with the difference in pion-quark couplings (from vertex  
renormalizations), we therefore conclude that $g_{\pi^0 nn}$ is numerically 
bigger than $g_{\pi^0 pp}$ by $0.80 \%$, while $g_{\pi^+ np}$ is numerically
greater than $g_{\pi^0 pp}$ by $0.15 \%$. 

Numerically, we may arbitrarily set $m_d = m_u$, $<{\bar d} d> =
<{\bar u} u>$, and $\lambda_n = \lambda_p$, except keeping physical
values for $m_n$ and $m_p$. In this way, we find that $\mid g_{\pi nn}
\mid$ is bigger than $\mid g_{\pi pp}\mid$ by 0.19 \%. Analogously,
we find that $\lambda_n \not= \lambda_p$ (as found by Yang et al.
\cite{Yang}) has a contribution of 0.20 \%. $m_d \not= m_u$ and
$<{\bar d} d> \not= <{\bar u} u>$ combine to give another contribution
of 0.20 \%. Therefore, the overall contribution of 80 \% in fact comes
from several sources of comparable magnitudes (and of the same sign in
the present case).

It is of some interest to note that our prediction on $\mid g_{\pi^0 nn}
\mid - \mid g_{\pi^0 pp}\mid $ (a positive value) differs in sign from 
the quark-model result \cite{Cao}, and also from a three-point correlation
function approach \cite{Meissner}. It is also of interest to note that
the difference in pion-quark couplings represents a source of numerical
importance on the charge symmetry breaking effect. 

\vskip 1.5 true cm
\section*{III. Low-Energy Nucleon-Nucleon Scattering}
\bigskip
The one-pion-exchange potential (OPEP) in nucleon-nucleon interactions
can be obtained in the standard manner \cite{Cao}. For example. we have,
in the $p-p$ case,
\begin{equation}
V_{pp}^\pi(r) = {g_{\pi^0 pp}^2\over 4\pi}\, { {\vec \sigma}\,^1 \cdot
{\vec \nabla}_r {\vec \sigma}\,^2 \cdot {\vec \nabla}_r \over 4 m_p^2}\, 
{1\over r}(e^{-m_{\pi^0}\, r} - e^{-\Lambda r}), 
\end{equation}
where $\Lambda$ is the cutoff at the short range which contributes negligibly
to the isospin symmetry breakings \cite{Cao}. The OPEP in the $n-n$ case
may be obtained by simple substitution $p \to n$. On the other hand, the 
$n-p$ OPE potential is given by
\begin{equation}
\begin{array}{ll}
V_{np}^\pi(r) = &{g_{\pi^0 nn}g_{\pi^0 pp}\over 4\pi}\, \tau_3^1 \tau_3^2
{ {\vec \sigma}\,^1 \cdot
{\vec \nabla}_r {\vec \sigma}\,^2 \cdot {\vec \nabla}_r \over 4 m_n m_p}\, 
{1\over r}(e^{-m_{\pi^0}\, r} - e^{-\Lambda r}) \\ 
& +  {g_{\pi^+ np}^2\over 4\pi}\, \{ ({\vec \tau}\,^1\cdot {\vec \tau}\,^2 
 -\tau_3^1\tau_3^2) { {\vec \sigma}\,^1 \cdot
{\vec \nabla}_r {\vec \sigma}\,^2 \cdot {\vec \nabla}_r \over 4 M_N^2}\\
&\qquad\qquad + {2\delta\over 4 M_N^2} ({\vec \tau}\,^1\times 
{\vec \tau}\,^2)_3
{\vec \sigma}\,^1\times{\vec \sigma}\,^2 \cdot {\vec L} {1\over r}{d\over dr}
\}\, {1\over r}(e^{-m_{\pi^+}\, r} - e^{-\Lambda r}). 
\end{array}
\end{equation}
Here we have used $\delta \equiv (m_n-m_p)/(m_n +m_p)$ and $M_N = {1\over 2}
(m_n +m_p)$. The $^1S_0$ phase shift $\delta_0$, as used to describe 
low-energy nucleon-nucleon scatterings, can be parametrized in the 
standard manner:
\begin{equation}
k cot\,\delta_0 = -{1\over a} + {1\over 2} k^2 r_0 + \ddots, 
\end{equation}
with $a$ the scattering length and $r_0$ the effective range. Experimentally,
we have \cite{Miller} 
\begin{equation}
\begin{array}{ll}
\delta a_{\mbox{CSB}} &\equiv \mid a_{pp}\mid - \mid a_{nn}\mid = -(1.5\pm 0.5) fm; \\
\delta a_{\mbox{CD}} &\equiv \mid a_{np}\mid - \mid a_{nn}\mid = (5.0\pm 0.3) fm.
\end{array}
\end{equation}
As already noted in \cite{Cao}, the differences in OPEP's, such as 
$V_{pp}(r) -V_{nn}(r)$, may be treated as a perturbation, making it easy 
to compute $\delta a_{\mbox{CSB}}$ or other low-energy isospin symmetry 
breaking observables.
In this way, our prediction on the difference between $g_{\pi^0 nn}$ 
and $g_{\pi^0 pp}$ gives 
rise to about $- 1.0\, fm$ on $\delta a_{\mbox{CSB}}$, 
consistent with the obervation both in sign and roughly in magnitude.

As already known \cite{Henley,Cao} for some time, the bulk of the 
charge-dependent effect as revealed by a (relatively) large value of
$\delta a_{\mbox{CD}}$ can be understood quantitatively as caused by the
pion mass difference $m_{\pi^+} \not= m_{\pi^0}$ while the differences
as found in pion-nucleon couplings give rise to relatively small 
contribution to $\delta a_{\mbox{CD}}$. As a result, we may safely conclude
that the observed $\delta a_{\mbox{CSB}}$ and $\delta a_{\mbox{CD}}$ as 
summarized recently in \cite{Miller} can be understood, to a large extent,
as manifestations of isospin asymmetries associated with the one-pion
exchange potential.

\vskip 1.5 true cm
\section*{IV. Discussion and Summary}
\bigskip
The observed differences \cite{Miller} in the $^1S_0$ $NN$ scatterings are 
by far the best known isospin symmetry breakings in which pion-nucleon
couplings play a central role. There might be some isospin symmetry 
breakings in the $^1S_0$ effective ranges or in the $^3P_J$ scattering
lengths, but until now such effects have not been extracted experimentally.

The last term in $V_{np}^\pi(r)$ [in Eq. (17)], as induced by $m_p \not=
m_n$, gives rise to isospin 
mixings such as $^3P_1 - ^1P_1$ or $^3D_2 - ^1D_2$ mixing, but only in
a rather small way. On the other hand, charge-symmetry breakings 
observed in medium-energy elastic $n-p$ scattering (at TRIUMF and IUCF) 
are dictated primarily \cite{Gardner} by the charge symmetry breaking
force as caused by the $\rho-\omega$ mixing or by $g_{\omega pp} \not=
g_{\omega nn}$. However, the short-range nature in this case may require
a careful investigation of the quark-interchange mechanisms as
addressed in \cite{Hwang2}, as it is difficult to reconcile the picture
of exchanging, between two nucleons, a meson ($\rho$ or $\omega$) with 
a range of about 0.25 fermi, considerably smaller than the extent of 
a nucleon (with a radius of at least 0.5 fermi, even assuming a little
bag picture for the nucleon). 

In any event, the isospin symmetry breakings associated with the 
one-pion exchange potential (which is clearly on a very sound conceptual
ground) can best be accessed through careful measurements of low-energy 
scattering parameters
such as $\delta a_{\mbox{CSB}}$, $\delta a_{\mbox{CD}}$, etc. Such efforts
should be further encouraged by the currently popular wisdom that the
low-energy sector of the $\pi N$ are $NN$ systems can be described by
the long-wave length limit of QCD via Goldstone theorem, yielding the 
so-called chiral perturbation theory, so that the low-energy scattering 
parameters seems to be more ``fundamental'' than those needed in, e.g., 
the description of medium-energy nucleon-nucleon scatterings.

In summary, we have adopted the method of QCD sum rules in the presence
of an external pion field to investigate isospin symmetry breakings
in pion-nucleon couplings. Numerically, we find that $\mid g_{\pi nn}
\mid $ is bigger than $\mid g_{\pi pp}\mid $ by $0.80\%$ (and bigger
than $\mid g_{\pi^+ np}\mid$ by $0.65\%$). This result accounts for 
the observed charge-symmtry breaking effect $\delta a_{\mbox{CSB}}$ both in
sign and (roughly) in magnitude. Contrary to many previous calculations
in the literature, our result offers, for the first time, the possibility
to understand both $\delta a_{\mbox{CSB}}$ and $\delta a_{\mbox{CD}}$ in a 
quantitative manner (both in sign and in magnitude).

\newpage
\begin{center}
{\bf Figure Captions}
\end{center}
\noindent
Fig. 1. Pictorial representation of the various terms in the quark
propagator $S^{(0)ab}(x)$.
\vskip 1cm
\noindent
Fig. 2. Pictorial representation of the various terms in the quark
propagator $S^{(\pi)ab}(x)$.
\vskip 1cm
\noindent
Fig. 3. The diagrams used to determine the QCD sum rule for $g_{\pi^0 pp}$.
\vskip 1cm
\noindent
Fig. 4. The vertex renormalizations (including the self-energy contributions) 
leading to the non-universal pion-quark couplings, an important source for
isospin symmetry breakings.

\end{document}